\documentclass[aps,showpacs,epsfig,twocolumn]{revtex4}
\usepackage{epsfig}

\newcommand{\be}{\begin{equation}}
\newcommand{\ee}{\end{equation}}
\newcommand{\bea}{\begin{eqnarray}}
\newcommand{\eea}{\end{eqnarray}}
\begin{document}

\title{\bf\Large {Canted Ferromagnetism in Double Exchange Model with on-site Coulomb
Repulsion}}

\author{Naoum Karchev and Vasil Michev\cite{byline}}

\affiliation{Department of Physics, University of Sofia, 1164 Sofia,
Bulgaria}

\begin{abstract}
The double exchange model with on-site Coulomb repulsion is
considered. Schwinger-bosons representation of the localized spins
is used and two spin-singlet Fermion operators are introduced. In
terms of the new Fermi fields the on-site Hund's interaction is in a
diagonal form and the true magnons of the system are identified. The
singlet fermions can be understood as electrons dressed by a cloud
of repeatedly emitted and reabsorbed magnons. Rewritten in terms of
Schwinger-bosons and spin-singlet Fermions the theory is $U(1)$
gauge invariant. We show that spontaneous breakdown of the gauge
symmetry leads to \emph{\textbf{canted ferromagnetism with on-site
spins of localized and delocalized electrons misaligned}}. On-site
canted phase emerges in double exchange model when Coulomb repulsion
is large enough. The quantum phase transition between ferromagnetism
and canted phase is studied varying the Coulomb repulsion for
different values of parameters in the theory such as Hund's coupling
and chemical potential.

\end{abstract}


\maketitle

\section {\bf Introduction}

Spin-fermion model describes materials which get their magnetic
properties from a system of localized magnetic moments being coupled
to conducting electrons. The model is known as $s-d$ (or $s-f$)
model in which the electrons are separated into delocalized $s$
electrons and localized $d$($f$) electrons. The names of the $s$ and
$d$($f$) electrons do not necessarily mean that the orbital electron
states are of corresponding type. They are introduced to distinguish
the localized from delocalized electrons. The model appears in the
literature also as the Ferromagnetic Kondo Lattice model (FKLM) or
the Double Exchange model
(DEM)\cite{cfm01,cfm02,cfm021,cfm03,cfm04,cfm05}.

The double exchange model, as we shall call the model from now on,
has great variety of applications to different topics in the
magnetism. On the basis of the double exchange model a microscopic
theory for ferromagnetic hexaborides, such as $EuB_6$ , is proposed
\cite{cfm1}. The magnetism of ferromagnetic metal is found to arise
from the half-filled $4f$ shell of $Eu$ , whose localized electrons
account for the measured moment \cite{cfm11,cfm12,cfm13}. The
transport properties such as the Hall effect, magnetoresistance, and
dc resistivity are quantitatively described within DEM \cite{cfm1}.

Another materials for which the DEM is relevant are the dilute
magnetic semiconductors, such as $Ga_{1-x}Mn_xAs$, where fraction of
non-magnetic elements $Ga$ is replaced with magnetic ones $Mn$. The
atoms of manganese supply of both carriers (holes) and magnetic
moment.These materials have attracted great attention after the
experimental observation of ferromagnetic transition temperature
$T_c$ as high as $110K$ \cite{cfm20}. One-band and two-band double
exchange models have been solved, by means of dynamical mean-field
approximation or Monte Carlo simulations, to find the magnetic
transition temperature as a function of coupling constants, hopping
parameters, and carrier densities \cite{cfm21,cfm22}.

The double exchange model is a widely used model for manganites
\cite{cfm01,cfm2}. In isolation, the ions of Mn  have an active
$3d$-shell with five degenerate levels. The degeneracy is presented
due to rotational invariance within angular momentum $l=2$ subspace.
The crystal environment results in a particular splitting of the
five $d$-orbitals  (\emph{crystal field spliting})into two groups:
the $\emph{e}_g$ and $\emph{t}_{2g}$ states. The electrons from the
$\emph{e}_g$ sector, which form a doublet, are removed upon hole
doping. The $\emph{t}_{2g}$ electrons, which form a triplet, are not
affected by doping, and their population remains constant. The Hund
rule enforces alignment of the three $\emph{t}_{2g}$ spins into a
$s=3/2$ state. Then, the $\emph{t}_{2g}$ sector can be replaced by a
\emph{localized spin} at each manganese ion, reducing the complexity
of the original five orbital model. The next drastic simplification
is that only one $\emph{e}_g$ orbital is available at each site. To
justify this one can assume \cite{cfm2} that a static Jahn-Teller
distortion leads to a splitting of the degenerate $\emph{e}_g$
levels, allowing to keep only one active orbital. The only important
interaction between the two sectors is the Hund coupling between
localized $\emph{t}_{2g}$ spins and mobile $\emph{e}_g$ electrons.

The double exchange model has a rich phase diagram, exhibiting a
variety of phases, with unusual ordering in the ground states. The
procedures followed to obtain the phase diagram are different:
numerical studies \cite{cfm31}, dynamical mean field theory
\cite{cfm32}, and analytical calculations \cite{cfm33,cfm34},  but
four phases have been systematically observed: (i)
antiferromagnetism (AF) at a density of mobile electrons $n=1$, (ii)
ferromagnetism (FM) at intermediate electronic densities, (iii)
phase separation (PS) between FM and AF phases, and (iv) spin
incommensurable (IC) phase at large enough Hund coupling. The
competition between spin spiral incommensurate order or phase
separation and canted ferromagnetism is also a topic of intensive
study \cite{cfm33,cfm34,cfm35}.

The simplest but realistic Hamiltonian for the double exchange model
has the form \be H =  - t\sum\limits_{  \langle  ij  \rangle  }
{\left( {c_{i\sigma }^ + c_{j\sigma } + h.c.} \right)}
-J_H\sum\limits_i {{\bf S}_i}\cdot {\bf s}_i \label{cfm1} \ee
 where $c_{i\sigma }^ +$ and $c_{i\sigma }$ are
creation and destruction operators for mobile electrons,
$s^{\mu}_i=\frac
12 \sum\limits_{\sigma\sigma'}c^+_{i\sigma}
\tau^{\mu}_{\sigma\sigma'}c^{\phantom +}_{i\sigma'}$, with the Pauli
matrices $(\tau^x,\tau^y,\tau^z)$, is the spin of the conduction
electrons, and ${\bf S}_i$ is the spin of the localized electrons.
The sums are over all sites of a three-dimensional cubic lattice,
and $\langle i,j\rangle$ denotes the sum over the nearest neighbors.
In equation (\ref{cfm1}) the hopping amplitude and the Hund coupling
between localized and mobile electrons are positive.

The Hamiltonian (\ref{cfm1}) of the DEM is quadratic with respect to
the fermions $(c_{i\sigma }^ +,c_{i\sigma })$. Averaging in the
subspace of the itinerant electrons one obtains an effective
Heisenberg like model in terms of core spins ${\bf S}_i$
\cite{cfm41,cfm42}. In the small $J_H$ limit
Ruderman-Kittel-Kasuya-Yosida (RKKY) theory is recovered. The subtle
point is that if we use a Holstein-Primakoff representation for the
localized spins ${\bf S}_i$, the creation and annihilation bose
operators do not describe the true magnon of the system \cite
{cfm43}. The true magnons are transversal fluctuations corresponding
to the total magnetization which includes both the spins of
localized and delocalized electrons. Therefore the RKKY validity
condition requires not only small Hund's coupling, but it also
insists the charge carrier density to be small, which in turn means
that the magnetization of the mobile electrons is inessential.

Since the only interaction between localized and delocalized
electrons is the Hund coupling, it is desirable to treat the on-site
term in the Hamiltonian (\ref{cfm1}) exactly. To this end, the
Holstein-Primakoff transformation was generalized to the case when
the length of the spin is operator by itself \cite{cfm51}. The
procedure removes all spin variables from the Hund coupling term.
The bosonic and fermionic sectors are constructed in $\frac {1}{s}$
expansion up to the fourth order. Alternatively, two spin-singlet
fermi fields are introduced in Ref. \cite{cfm52}. In terms of the
singlet Fermi fields the on-site term is in a diagonal form, the
spin variables are removed, and one can treat it exactly. An
analogous technique is used in Ref. \cite{cfm53}.

More realistic DEM would account for the on-site Coulomb repulsion.
The Coulomb (Hubbard) interaction has a profound effect on the band
structure, magnetic ground state (magnetic configurations), and
transport properties of the spin-fermion systems. The results of the
electron-electron repulsion depend on parameters in the theory such
as doping, band width, and temperature. While some effects of the
Hubbard term have been addressed in the past by means of mean-field
theory \cite{cfm61,cfm62,cfm63}, local spin-density approximation
(LSDA) and LSDA+U calculations \cite{cfm64}, and dynamical
mean-field theory \cite{cfm65} its impact is not fully appreciated
so far.

\begin{figure}[!ht]
\vspace{0.02cm} \epsfxsize=5cm \hspace*{0.2cm}
\epsfbox{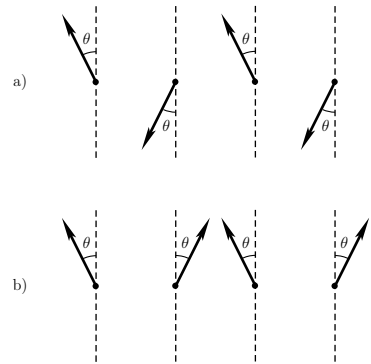} \\ \hskip 1cm \caption{Sketch of two
sublattice spin-canted states:  \\a) canted antiferromagnetism,\quad
b) canted ferromagnetism } \label{fig1}
\end{figure}

In the present paper we study canted ferromagnetism in double
exchange model with on-site Coulomb repulsion. Usually the canted
magnetism is considered as a two-sublattice spin configuration with
neighboring lattice spins misaligned by an angle $ \theta $. First
de Gennes observed that in double exchange model spin-canted state
(Fig. \ref{fig1}a) interpolates between ferromagnetic and
antiferromagnetic order \cite{cfm71}. The canted phase (Fig.
\ref{fig1}b) is a part of the phase diagram of mixed-spin $(S_1
\rangle S_2)$ $J_1-J_2$ Heisenberg model on square lattice
\cite{cfm72}. Finally, canted phase appears in the lattice models of
quantum rotors \cite{cfm73}.

\begin{figure}[!ht]
\vspace{0.02cm} \epsfxsize=4.5cm \hspace*{0.2cm}
\epsfbox{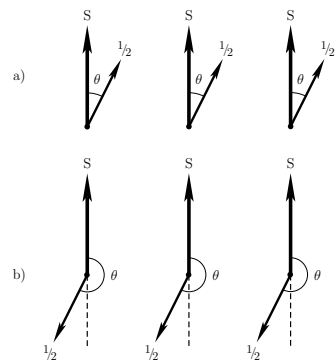} \\
\caption{Sketch of on-site spin-canted states: \\ a) when $J_H >
0$,\quad b) when $J_H < 0$} \label{fig2}
\end{figure}

In the present paper we show that \emph{\textbf{canted
ferromagnetism with on-site spins of localized and delocalized
electrons misaligned}} (Fig. \ref{fig2}) emerges in double exchange
model when Coulomb repulsion is large enough.
We study quantum phase transition between ferromagnetic and on-site
canted orders when Coulomb repulsion is varied for different values
of parameters in the theory such as Hund's coupling and chemical
potential.

The paper is organized as follows. In Sec.II we study double
exchange model (\ref{cfm1}) supplemented with an antiferromagnetic
Heisenberg interaction between nearest-neighbors core spins.
Schwinger-bosons representation of the localized spins is used and
two spin-singlet Fermion operators are introduced. In terms of the
new Fermi fields the on-site Hund's interaction is in a diagonal
form and the true magnons of the system are identified. The singlet
fermions can be understood as electrons dressed by a cloud of
repeatedly emitted and reabsorbed magnons. Integrated over the
singlet fermions we obtain an effective Heisenberg-like theory.
Positivity of the spin-stiffness, as a function of Hund's coupling
$J_H$ and charge carrier density, is the condition for the stable
ferromagnetism. The phase boundary is depicted both in the case of
zero and non-zero antiferromagnetic exchange. Sec. III is devoted to
the on-site canted ferromagnetism in DEM with on-site Coulomb
repulsion. The theory rewritten in terms of Schwinger-bosons and
spin-singlet Fermions is an $U(1)$ gauge invariant theory. We show
that on-site canted state is a state with spontaneously broken gauge
symmetry. The quantum phase transition between ferromagnetism and
canted phase is studied varying the parameters in theory. A summary
in Sec. IV concludes the paper.

\section{\bf Magnons in Double Exchange Model}

We consider a theory with Hamiltonian \bea \nonumber h & = & H-\mu N
= -t\sum\limits_{  \langle  ij
 \rangle  } {\left( {c_{i\sigma }^ + c_{j\sigma } + h.c.} \right)}  -\mu \sum\limits_i {n_i
} \\ & + & J_{AF}\sum\limits_{  \langle  ij  \rangle  } {{\bf S}_i
\cdot {\bf S}_j}
  - J_H\sum\limits_i {{\bf
S}_i}\cdot {\bf s}_i \label{cfm2}\eea where $\mu$ is the chemical
potential, and $n_i=c^+_{i\sigma}c_{i\sigma}$. The antiferromagnetic
Heisenberg term $(J_{AF} \rangle 0)$ is very important for the
manganites. In the limit when all $e_{g}$ electrons are removed, and
the system is without mobile electrons, the $t_{2g}$ electrons
induce an antiferromagnetic Heisenberg exchange between
nearest-neighbors leading to the standard antiferromagnetism. The
most prominent example is $CaMnO_3$ \cite{cfm2}.

In terms of Schwinger-bosons ($\varphi_{i,\sigma},
\varphi_{i,\sigma}^{\dagger}$) the spin operators have the following
representation: \be {\bf S}_i = \frac{1}{2}\varphi _{i\sigma }^ +
{\bf \tau} _{\sigma \sigma '} \varphi _{i\sigma '}, \qquad \varphi
_{i\sigma }^ + \varphi _{i\sigma }  = 2s. \label{cfm3}\ee The
partition function can be written as a path integral over the
complex functions of the Matsubara time
$\varphi_{i\sigma}(\tau)$\,\,$(\varphi_{i\sigma}^+(\tau))$ and
Grassmann functions $c_{i\sigma }(\tau)$\,\,$(c^+_{i\sigma}(\tau))$.
\be {\cal
Z}(\beta)\,=\,\int\,d\mu\left(\varphi^+,\varphi,c^+,c\right) e^{-S}.
\label{cfm4} \ee with an action given by the expression \bea
\nonumber S & = & \int\limits^{\beta}_0 d\tau\left[
\sum\limits_i\left(\varphi^+_{i\sigma} (\tau)
\dot\varphi_{i\sigma}(\tau)+c^+_i(\tau)\dot c_i(\tau)\right) +
\right.
\\ & & \left.  h\left(\varphi^+,\varphi,c^+,c\right)\,\right], \label{cfm5}
\eea where $\beta$ is the inverse temperature and the Hamiltonian is
obtained from Eqs.(\ref{cfm2}) and (\ref{cfm3}) replacing the
operators with the functions. In terms of Schwinger-bosons the
theory is invariant under $U(1)$ gauge transformations
\be\label{cfm6} \varphi'_{j\sigma}(\tau)=
e^{i\alpha_j(\tau)}\varphi_{j\sigma}(\tau);\,\,
\varphi'^+_{j\sigma}(\tau)=
e^{-i\alpha_j(\tau)}\varphi^+_{j\sigma}(\tau) \ee with parameters
which are period functions of Matsubara time
$\alpha_j(0)=\alpha_j(\beta)$. The measure for the Schwinger bosons
includes Dirac-$\delta$ functions that enforce the constraint
(\ref{cfm3}) and the gauge-fixing condition \bea
D\mu\left(\varphi^+,\varphi\right) & = &
 \prod\limits_{i,\tau,\sigma}\frac {D\varphi^+_{i\sigma}(\tau)
D\varphi_{i\sigma}(\tau)}{2\pi i} \\
& & \prod\limits_{i\tau}\delta\left(\varphi^+_{i\sigma}(\tau)
\varphi_{i\sigma}(\tau)-2s\right)
\prod\limits_{i\tau}\delta\left(g.f\right). \nonumber\label{mp6}
\eea

We introduce two spin-singlet Fermi fields  \bea & &
\Psi^A_i(\tau)=\frac {1}{\sqrt
{2s}}\varphi^+_{i\sigma}(\tau)c_{i\sigma}(\tau),\label{cfm8}
\\ & &  \Psi^B_i(\tau)=\frac {1}{\sqrt
{2s}}\left[\varphi_{i1}(\tau)c_{i2}(\tau)\,-
\,\varphi_{i2}(\tau)c_{i1}(\tau)\right], \label{cfm9} \eea which are
gauge variant with charge -1 and 1 respectively \be \label{cfm10}
\Psi'^A_j(\tau)=e^{-i\alpha_j(\tau)}\Psi^A_j(\tau),\,\,\,\,
\Psi'^B_j(\tau)=e^{i\alpha_j(\tau)}\Psi^B_j(\tau).\ee The equations
(\ref{cfm8}) and (\ref{cfm9}) can be regarded as a SU(2)
transformation \be \Psi_{i\sigma}  =
g_{i\sigma\sigma'}^+c_{i\sigma'}\,\, \Rightarrow \,\, g_i^+   =
\frac{1}{{\sqrt {2s} }}\left( {\begin{array}{*{20}c}
   {\varphi_{i1}^ +  } & {\varphi_{i2}^ +  }  \\
   { - \varphi_{i2} } & {\varphi_{i1} }  \\
\end{array}} \right)\label{cfm11}\ee
with $\Psi^A_i=\Psi_{1i}$ and $\Psi^B_i=\Psi_{2i}$.
 For that reason the Fermi measure is invariant under the
change of variables. In terms of the spin-singlet Fermi fields the
spin of the conduction electrons ${\bf s}_i$ has the form
\be\label{cfm12} s_i^{\mu} = \frac 12 c^+_{i\sigma}
\tau^{\mu}_{\sigma\sigma'}c^{\phantom +}_{i\sigma'} = \frac 12{\rm
O_i}^{\mu \nu }\Psi_{i\sigma}^+
\tau^{\nu}_{\sigma\sigma'}\Psi_{i\sigma'}, \ee where
\be\label{cfm13} {\rm O_i}^{\mu \nu }  = \frac 12\mathop{Tr} g_i^ +
\tau ^\mu g_i\tau ^\nu.\ee It is convenient to introduce three basic
vectors which depend on the Schwinger-bosons \be\label{cfm14}
T^1_{i\mu}= {\rm O_i}^{\mu 1}\quad T^2_{i\mu}= {\rm O_i}^{\mu
2}\quad T^3_{i\mu}= {\rm O_i}^{\mu 3},\ee where ${\bf T}^3_{i}=
\frac 1s {\bf S}_i$. Then, the spin of the electrons can be
represented as a linear combination of three vectors ${\bf S}_j$,
${\bf P}_j={\bf T}^1_j+i{\bf T}^2_j$ and ${\bf P}^+_j={\bf
T}^1_j-i{\bf T}^2_j$ \bea\label{cfm15} {\bf s}_i & = &
\frac{1}{{2s}}{\bf
S}_i\left( {\Psi^{A+}_i\Psi^A_i  - \Psi^{B+}_i \Psi^B_i} \right)\ \\
& + & \frac{1}{2} {\bf P}_i \Psi^{B+}_i\Psi^A_i + \frac{1}{2}{\bf
P}^+_i \Psi^{A+}_i\Psi^B_i.  \nonumber \eea The basic vectors
satisfy the relations ${\bf S}_i^2=s^2$, \,\,${\bf P}^2_i={\bf
P}^{+2}_i= {\bf S}_i\cdot{\bf P}_i= {\bf S}_i\cdot{\bf P}^+_i=0$,
and ${\bf P}^+_i\cdot{\bf P}_i=2$. Using the expression
(\ref{cfm15}) for the spin of itinerant electrons the total spin of
the system ${\bf S}^{tot}_i={\bf S}_i+{\bf s}_i$ can be written in
the form \bea {\bf S}^{tot}_i & = & \frac 1s \left[s+\frac 12
\left (\Psi^{A+}_i\Psi^A_i-\Psi^{B+}_i\Psi^B_i\right)\right]{\bf S}_i\,+ \nonumber \\
& & \frac{1}{2} {\bf P}_i \Psi^{B+}_i\Psi^A_i + \frac{1}{2}{\bf
P}^+_i \Psi^{A+}_i\Psi^B_i \label{cfm16} \eea

The gauge invariance imposes the conditions \\
$\langle\Psi^{A+}_i\Psi^B_i\rangle=  \langle \Psi^{B+}_i\Psi^A_i
\rangle =0$. As a result, the dimensionless magnetization per
lattice site $M= \langle (S^{tot}_i)^z \rangle $ reads
\begin{equation}
M=\frac 1s \left[s+\frac 12  \langle \left
(\Psi^{A+}_i\Psi^A_i-\Psi^{B+}_i\Psi^B_i\right) \rangle \right]
\langle {\bf S}^{z}_i \rangle  \label{cfm17}
\end{equation}
At zero temperature $ \langle {\bf S}^{z}_i \rangle =s$ and $M=s+m$,
where
\begin{equation}
m=\frac 12 \langle \left
(\Psi^{A+}_i\Psi^A_i-\Psi^{B+}_i\Psi^B_i\right) \rangle
\label{cfm18}
\end{equation}
is the contribution of the itinerant electrons.

Let us average  the total spin of the system (Eq. \ref{cfm16}) in
the subspace of the itinerant electrons $  \langle {\bf S}^{tot}_i
\rangle _f = {\bf M}_i$. The vector ${\bf M}_i$ $({\bf M}_i^2=M^2)$
identifies the local orientation of the total magnetization.
Accounting for the gauge invariance, one obtains an expression for
${\bf M}_i$ in terms of core spins ${\bf S}_i$ \be\label{cfm19}
\langle {\bf S}^{tot}_i \rangle _f = {\bf M}_i = \frac Ms {\bf S}_i
\ee Now, if we use Holstein-Primakoff representation for the vectors
${\bf M}_j$ \bea\label{cfm20} & &
M_j^+ = M_{j1} + i M_{j2}=\sqrt {2M-a^+_ja_j}\,\,\,\,a_j \nonumber \\
& & M_j^- = M_{j1} - i M_{j2}=a^+_j\,\,\sqrt {2M-a^+_ja_j}
\\ & & M^3_j = 2M - a^+_ja_j \nonumber \eea the bose fields
$a_j$ and $a^+_j$ are the \textbf{true magnons} in the system. In
terms of the true magnons the Schwinger-bosons (\ref{cfm3}) have the
following representation \be\label{cfm21} \varphi_{i1} = \sqrt
{2s-\frac sM a^+_ia_i}, \qquad \varphi_{i2} = \sqrt {\frac sM}\,\,\,
a_i\,\,. \ee Replacing in Eqs. (\ref{cfm8}) and (\ref{cfm9}) for the
spin-singlet Fermions and keeping only the first two terms in $1/M$
expansion $\sqrt{1-\frac {1}{2M}\,\,a^+_ia_i}\simeq 1-\frac
{1}{4M}\,\,a^+_ia_i +...$ \,\,we obtain \bea & & \Psi^A_i = c_{i1} +
\frac {1}{\sqrt{2M}}\,\,a^+_ic_{i2} - \frac {1}{4M}
a^+_ia_ic_{i1}+....\,\,,\label{cfm22}
\\ & &  \Psi^B_i =  c_{i2} -
\frac {1}{\sqrt{2M}}\,\,a_ic_{i1} - \frac {1}{4M}
a^+_ia_ic_{i2}+....\,\,.\label{cfm23}\eea The equations
(\ref{cfm22}) and (\ref{cfm23}) show that the singlet fermions are
electrons dressed by a virtual cloud of repeatedly emitted and
reabsorbed magnons.

An important advantage of working with $A$ and $B$ fermions is the
fact that in terms of these spin-singlet fields the spin-fermion
interaction is in a diagonal form, the spin variables (magnons) are
removed, and one accounts for it exactly
\be\label{cfm24}\sum\limits_{i}{{\bf S}}_i\cdot{{\bf s}}_i=\frac
s2\sum\limits_{i} [\Psi^{A+}_i\Psi^A_i-\Psi^{B+}_i\Psi^B_i]\,\,.\ee

To proceed we rewrite the action (\ref{cfm5}) as a function of
Schwinger-bosons and spin-singlet fermions
\begin{eqnarray} S & = & \int\limits_0^\beta  d\tau \Bigg[
\varphi_{i\sigma }^ +  \dot \varphi_{i\sigma } + \Psi^{A+}_{i}
\left( {\frac{\partial }{{\partial \tau }} +
\frac{1}{{2s}}\varphi_{i\sigma
}^ + \dot \varphi_{i\sigma } } \right) \Psi^A_{i} \nonumber \\
& + & \Psi _{i}^{B+} \left( {\frac{\partial }{{\partial \tau }} -
\frac{1}{{2s}}\varphi_{i\sigma }^ +  \dot \varphi_{i\sigma } }
\right)\Psi^B_{i} \nonumber \\
 & + & \frac{1}{{2s}}\left( { - \varphi_{i1}^ +
\dot \varphi_{i2}^ + + \varphi_{i2}^ + \dot \varphi_{i1}^ +}
\right)\Psi _{i}^{A+} \Psi^B_{i} \label{cfm25} \\
& + & \frac{1}{{2s}}\left( { - \varphi_{i2} \dot \varphi_{i1} +
\varphi_{i1} \dot \varphi_{i2} } \right)\Psi _{i}^{B+} \Psi^A_{i} \nonumber \\
& + & h\left( {\varphi^+ ,\varphi,\Psi^+ ,\Psi }\right)\Bigg]\,\,.
\nonumber \end{eqnarray} It is convenient to write the Hamiltonian
$h\left( {\varphi^+ ,\varphi,\Psi^+ ,\Psi }\right)$ as a sum of
three terms \be \label{cfm26} h = h_f + h_H + h_{int}\,\,, \ee where
$h_f$ is the free \textbf{A} and \textbf{B} fermions' Hamiltonian
\bea\label{cfm27} h_f & = & - t\sum\limits_{ \langle  ij  \rangle }
{\left ({\Psi _{i\sigma }^ + \Psi _{j\sigma } + h.c.} \right)}
 - \mu \sum\limits_i {\Psi _{i\sigma }^ + \Psi _{i\sigma } }
 \nonumber \\
& - & \frac{{J_H s}}{2}\sum\limits_i {\left( {\Psi_{i}^{A+}
\Psi^A_{i} - \Psi_{i}^{+B}\Psi^B_{i} } \right)}\,\,,
\eea $h_H$ is the Hamiltonian of Heisenberg theory of
antiferromagnetism (\ref{cfm2}), and $h_{int}$ is the Hamiltonian of
magnon-fermion interaction

\begin{eqnarray}\label{cfm28} & & h_{\rm {int}} =
 -  t\sum\limits_{  \langle  ij  \rangle  } \Bigg[\left[ \frac{1}{{2s}}\left(
{\varphi_{i\sigma }^ +  \varphi_{j\sigma }  - 2s}
\right)\Psi_{i\sigma'}^{+}\Psi_{j\sigma'}
 + h.c.\right] \nonumber \\
 & & \\
& + & \left[\frac{1}{{2s}}\left( {\varphi_{i1}^ +  \varphi_{j2}^ + -
\varphi_{j1}^ + \varphi_{i2}^ + } \right)\left( {\Psi _{j}^{A+}
\Psi^B_{i} - \Psi _{i}^{A+} \Psi^B _{j} } \right) +
h.c.\right]\Bigg]\nonumber
\end{eqnarray}

The action (\ref{cfm25}) is quadratic with respect to the
spin-singlet fermions and one can integrate them out.  We can
accomplish this by first using the representation (\ref{cfm21}) of
the Schwinger-bosons, then keeping only the quadratic terms with
respect to the magnons, and  finally calculating the diagrams in the
leading order of gradient expansion. The action of the effective
theory, in Gaussian approximation is \bea\label{cfm29} S_{\rm eff} &
= &
\int\limits_0^\beta  d\tau \Bigg[a_i^+ \dot a_i \\
& + & MJ \sum\limits_{  \langle  ij  \rangle  }\left(a_i^+a_i +
a_j^+a_j - a_i^+a_j - a_j^+a_i \right)\Bigg], \nonumber\eea where
$M$ is the dimensionless magnetization per lattice site Eq.
(\ref{cfm17}) at zero temperature, and $J$ is the effective exchange
coupling
\begin{eqnarray}\label{cfm30}
J & = & -\frac {s^2}{M^2} J_{AF}  \\
& + & \frac {t}{6M^2}
\int\limits_{-\pi}^{\pi}\int\limits_{-\pi}^{\pi}\int\limits_{-\pi}^{\pi}
\frac {d^3k}{(2\pi)^3}
\left(\sum\limits_{\mu=1}^3 \cos k_{\mu}\right)\left(n_k^A+n_k^B\right) \nonumber \\
& - & \frac {2t^2}{3M^2J_H
s}\int\limits_{-\pi}^{\pi}\int\limits_{-\pi}^{\pi}\int\limits_{-\pi}^{\pi}
\frac {d^3k}{(2\pi)^3}\left(\sum\limits_{\mu=1}^3\sin^2
k_{\mu}\right)\left(n^A_k-n^B_k\right) \nonumber
\end{eqnarray}
In equation (\ref{cfm30}) $n_k^R= \theta
\left(-\varepsilon^R_k\right)$ (R=A, B) are the occupation numbers
for the A and B fermions with dispersions
\begin{eqnarray}\label{cfm31}
 \varepsilon^A_k & = & -2t\left(\cos k_x + \cos k_y +\cos k_z \right) - \mu -\frac {J_H s}{2} \nonumber\\
 \\
\varepsilon^B_k & = & -2t\left(\cos k_x + \cos k_y +\cos k_z \right)
- \mu  + \frac {J_H s}{2}\nonumber
\end{eqnarray}
The first term in equation (\ref{cfm30}) comes from "tadpole"
diagrams with one A or B-fermion line with vertices which relate to
the first term in the Hamiltonian of interaction (\ref{cfm28}). The
second term is obtained calculating the one-loop diagrams with A and
B-fermion lines, and with vertices which relate to the second term
in $h_{int}$. The term with time derivative in the effective action
(\ref{cfm29}) is obtained summing two terms. The first one is the
term with time derivative in the action (\ref{cfm25}) which in terms
of magnons has the form $\int\limits_0^\beta  d\tau \frac
{s}{M}a^+_i \dot a_i$, while the second results from "tadpole"
diagrams with vertices related to the second and third terms of the
action (\ref{cfm25}).

Based on the rotational symmetry, one can supplement the action
(\ref{cfm29}) up to an effective Heisenberg like action, written in
terms of the vectors ${\bf M_i}$ \be\label{cfm32} H_{\rm eff}= -
J\sum\limits_{  \langle  ij  \rangle  }{\bf M_i}\cdot{\bf M_i}. \ee
The ferromagnetic phase is stable if the effective exchange coupling
constant is positive $J \rangle 0$. The dimensionless constant
$J/W$, where $W=12t$ is the band-width, depends on $J_{AF}/W, J_H
s/W$ and $\mu/2t$. The $(\frac {J_H s}{W},n)$ phase diagram, where
$n$ is carrier density, is depicted in Fig.3 for $J_{AF}=0$ and
$\frac {J_{AF}}{W}=0.1$.
\begin{figure}[!ht]
\begin{center}
\hskip -0.0cm \vspace{0.02cm} \epsfxsize=8.5cm
\epsfbox{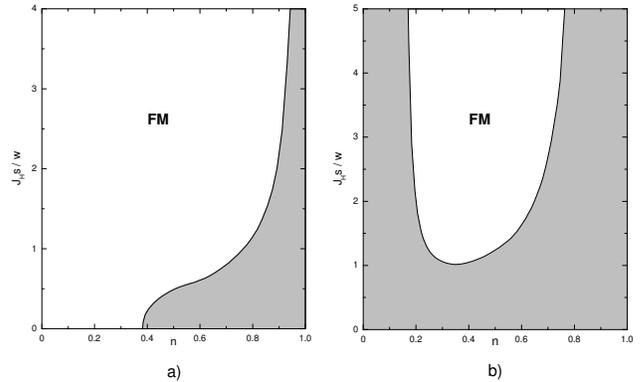} \caption{Phase diagrams when  a) $J_{AF}=0$
and b) $\frac {J_{AF}}{W}=0.1$} \label{fig3}
\end{center}
\end{figure}

The phase diagram Fig.3a $(J_{AF}=0)$ is in a good agreement with
phase diagrams obtained numerically \cite{cfm31} and by means of
alternative analytical calculations \cite{cfm34}. Phase diagram
Fig.3b shows that direct antiferromagnetic exchange suppresses the
ferromagnetism at small values of carrier concentrations, which is
well known experimental fact for manganites \cite{cfm2}.

\section{\bf Canted Ferromagnetism}

After considering pure double exchange model, let us address the
double exchange model supplemented with on-site Coulomb repulsion
(Hubbard term). \bea \nonumber h & = & H-\mu N = -t\sum\limits_{
\langle ij
 \rangle  } {\left( {c_{i\sigma }^ + c_{j\sigma } + h.c.} \right)} \\
& - & J_H\sum\limits_i {{\bf S}_i}\cdot {\bf s}_i  + U
\sum\limits_i n_{i\uparrow}n_{i\downarrow} -\mu \sum\limits_i {n_i
}\label{cfm33}, \eea where $n_{i\sigma}=c^+_{i\sigma}c_{i\sigma}$.
Our purpose is to show that canted ferromagnetism, with on-site
spins of localized and delocalized electrons misaligned, emerges
in double exchange model when Coulomb repulsion is large enough.

Let us average  the spin of the electrons (Eq. \ref{cfm15}) in the
subspace of the itinerant electrons. We obtain, as a consequence
of gauge invariance, that spin of electrons are parallel to the
localized spins $  \langle {\bf s}_i \rangle _f = \frac ms {\bf
S}_i$. Equation (\ref{cfm15}) shows that the on-site spins are
misaligned if $ \langle \Psi^{A+}_i\Psi^B_i \rangle $ and $
\langle \Psi^{B+}_i\Psi^A_i \rangle $ are not equal to zero, which
in turn means that gauge symmetry is spontaneously broken. To
explore dynamical breakdown of the gauge symmetry, or which is the
same, the on-site canted ferromagnetism we rewrite the Hamiltonian
(\ref{cfm33}) in terms of Schwinger-bosons and spin-singlet
Fermions. In particular, one obtains for the Hubbard term
\begin{equation}
\sum\limits_{i} n_{i\uparrow} n_{i\downarrow}= -
\sum\limits_{i}\Psi^{A+}_i\Psi^B_i\Psi^{B+}_i\Psi^A_i. \label{cfm34}
\end{equation}
We decouple this term by means of the Hubbard-Stratanovich
transformation, introducing complex field $\Delta_i$ $(\Delta^+_i)$,
the order parameter of the gauge symmetry breaking.
\begin{eqnarray}\label{cfm35}
& & e^{U\int\limits^{\beta}_0 d\tau\sum\limits_i
\Psi^{A+}_i(\tau)\Psi^B_i(\tau)\Psi^{B+}_i(\tau)\Psi^A_i(\tau)}
\\
& & =\int d\mu(\Delta^+\Delta) \exp \Bigg [-\int\limits^{\beta}_0
d\tau\sum\limits_i \Bigg
 [\frac {\Delta^+_i(\tau)\Delta_i(\tau)}{U} \nonumber \\
& + &  \Psi^{A+}_i(\tau)\Psi^B_i(\tau)\Delta_i(\tau)+
 \Delta^+_i(\tau)\Psi^{B+}_i(\tau)\Psi^A_i(\tau)\Bigg]\Bigg]
\nonumber
\end{eqnarray}
Now, the partition function (\ref{cfm4}) can be represented as a
path integral over the spin-singlet fermions, Schwinger-bosons, and
complex order parameter. The integral over the fermions is Gaussian,
and one can integrate them out. The resulting expression for the
partition function is an integral over the  Schwinger-bosons, and
complex order parameter. \be {\cal
Z}(\beta)\,=\,\int\,d\mu\left(\varphi^+,\varphi,\Delta^+,\Delta\right)
e^{-W\left(\varphi^+,\varphi,\Delta^+,\Delta \right) }.
\label{cfm36} \ee We perform the integral over the collective
variables $\Delta^+_i$ and $\Delta_i$ using the steepest descend
method. To this end, we set the spin fluctuation $a^+_i$ and $a_i$
(see equations (\ref{cfm20},\ref{cfm21})) equal to zero and assume
that mean-field value of the order parameter $\Delta_i(\tau)$ is an
independent of $\tau$ and lattice sites $i$ real constant $\Delta$.
Then, the free energy of the system, in mean-field approximation, is
\be\label{cfm37}  {\cal F} = \frac {\Delta^2}{U}- {\cal F}_f \ee
where ${\cal F}_f$ is the free energy of a Fermi system with
Hamiltonian \bea\label{cfm38} h_f & = & \sum\limits_k \bigg [
\varepsilon^A_k \Psi^{A+}_k\Psi^A_k + \varepsilon^B_k
\Psi^{B+}_k\Psi^B_k  \\
 & + & \Delta \left( \Psi^{A+}_k\Psi^B_k + \Psi^{B+}_k\Psi^A_k
 \right) \bigg ]\nonumber \eea To write the Hamiltonian in diagonal
 form one introduces new Fermi fields $\psi^a_k$ and $\psi^b_k$
 \be\label{cfm39} \Psi^A_k=\textit{u}\,\, \psi^a_k+\textit{v}\,\,\psi^b_k, \qquad
\Psi^B_k=- \textit{v}\,\, \psi^a_k+ \textit{u}\,\,\psi^b_k ,\ee
where the coefficients are \bea\label{cfm40} \textit{u} & = &
\sqrt{\frac 12\left(1+\frac
{J_H\,s}{\sqrt{(J_H\,s)^2+4\Delta^2}}\right)},\nonumber \\
\\
\textit{v} & = & (\rm sign \Delta) \sqrt{1- \textit{u}^2} \nonumber
\eea Then, \be\label{cfm41}h_f  =  \sum\limits_k \left [
\varepsilon^a_k \psi^{a+}_k\psi^a_k + \varepsilon^b_k
\psi^{b+}_k\psi^b_k \right ].\ee Here  $\varepsilon^{a}_k =
\varepsilon^{-}_k, \varepsilon^{b}_k = \varepsilon^{+}_k$, where
\be\label{cfm42} \varepsilon^{\pm}_k= \varepsilon_k \pm \frac {1}{2}
\sqrt{\left(J_H\,s\right)^2 +
 4\Delta^2},\ee
and $\varepsilon_k = -2t
\left[\cos(k_x)+\cos(k_y)+\cos(k_z)\right]-\mu $. Now, we can obtain
the mean-field expression for the free energy. At zero temperature
it is \be\label{cfm43}{\cal F} = \frac {\Delta^2}{U} -
\int\limits_{-\pi}^{\pi}\int\limits_{-\pi}^{\pi}\int\limits_{-\pi}^{\pi}
\frac {d^3k}{(2\pi)^3}\left[\varepsilon^{a}_k  \theta
\left(-\varepsilon^a_k\right) + \varepsilon^{b}_k  \theta
\left(-\varepsilon^b_k \right)\right] \ee

It is convenient to introduce the angle between the on-site spin of
carrier and localized spin \be\label{cfm44} \cos  \Theta  = \frac
{\textbf{S}_i\cdot\textbf{s}_i}{|\textbf{S}_i||\textbf{s}_i|}.\ee
Using the equation (\ref{cfm15}), for the spin of itinerant
electrons, and the properties of the basic vectors one obtains
\be\label{cfm45} \cos  \Theta  = \Bigg [ 1 + \frac {4
 \langle \Psi^{A+}_i\Psi^B_i \rangle  \langle  \Psi^{B+}_i\Psi^A_i \rangle }{ \langle \Psi^{A+}_i\Psi^A_i -
\Psi^{B+}_i\Psi^B_i \rangle } \Bigg ]^{-\frac 12} \ee We calculate
the matrix elements in the formula (\ref{cfm45}) in mean-field
approximation applying the transformation (\ref{cfm39},\ref{cfm40}).
The result is \bea\label{cfm46} & &  \langle \Psi^{A+}_i\Psi^A_i -
\Psi^{B+}_i\Psi^B_i \rangle  = \frac {J_H\,s\,\left(n^a -
n^b\right)}{\sqrt{\left(J_H\,s\right)^2 +
4\Delta^2}} \\
& &  \langle \Psi^{A+}_i\Psi^B_i \rangle  =  \langle
\Psi^{B+}_i\Psi^A_i \rangle  = \frac {\Delta\,\left(n^a -
n^b\right)}{\sqrt{\left(J_H\,s\right)^2 + 4\Delta^2}},\nonumber \eea
where $n^a$ and $n^b$ are occupation numbers for $"a"$ and $"b"$
fermions introduced by the transformation (\ref{cfm39}). After some
algebra we arrive at the mean-field expression for the angle
\be\label{cfm47} \cos  \Theta  = \frac
{J_H\,s}{\sqrt{\left(J_H\,s\right)^2 + 4\Delta^2}}. \ee Next, we
replace $\Delta$ in equations (\ref{cfm42}) and (\ref{cfm43}) by
$\cos \Theta $ from (Eq.\ref{cfm47}) and rewrite the mean-field free
energy as a function of $\cos \Theta $. The dimensionless energy $F
= 6{\cal F}/W$ is depicted in (Figs .4, 5, 6) for different values
of $W/U$ and fixed $J_H\,s/W$ and $\mu/W$. As graphs show,
increasing the Coulomb repulsion constant the system passes trough a
first order quantum phase transition. Red lines correspond to the
critical values $U_c$ of the Coulomb repulsion. The values $U_c$ and
$ \Theta _c$ depend on the parameters of the theory such as Hund's
coupling constant, chemical potential and band width . The character
of the transition also depends on the parameters in the theory. We
see (Fig. 6) that when $J_H\,s/W=0.50$ and $\mu/W=-0.33$ for small
values of Coulomb repulsion, $W/U$=1.50, 1.10, the minimum of the
free energy is at $\cos \Theta =1$. Near the quantum phase
transition $W/U=0.77$ the ground state is highly degenerated, while
below this critical value, for large enough $U$, the on-site canted
ferromagnetic state, with $\cos \Theta <1$, is the ground state of
the system. To figure that quantum phase transition out one has to
go beyond the mean-field theory or to use alternative methods of
calculations to complement the above one.
\begin{figure}[!ht]
\begin{center}
\vspace{-0.cm} \epsfxsize=7cm \hspace*{0.2cm} \epsfbox{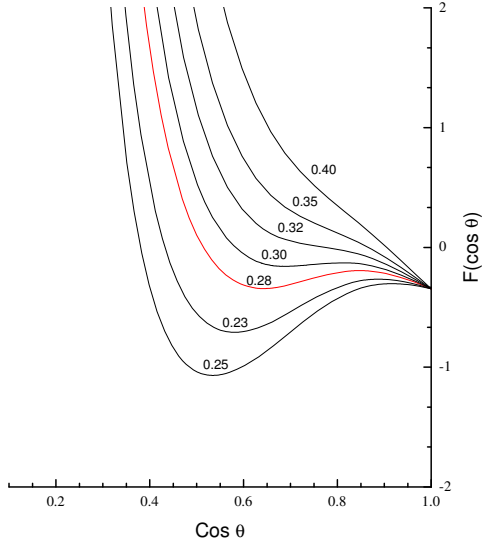}
\vskip -.5cm\caption{Dimensionless mean-field free energy $F=6{\cal
F}/W$ as a function of $\cos \theta $ for $J_H\,s/W =2.32$, $\mu/W
=-1.22$, and $W/U$=0.23; 0.25; 0.28; 0.30; 0.32; 0.35;
0.40}\label{fig4}
\end{center}
\end{figure}

\begin{figure}[!ht]
\begin{center}
\vspace{-0.cm} \epsfxsize=7cm \hspace*{0.2cm} \epsfbox{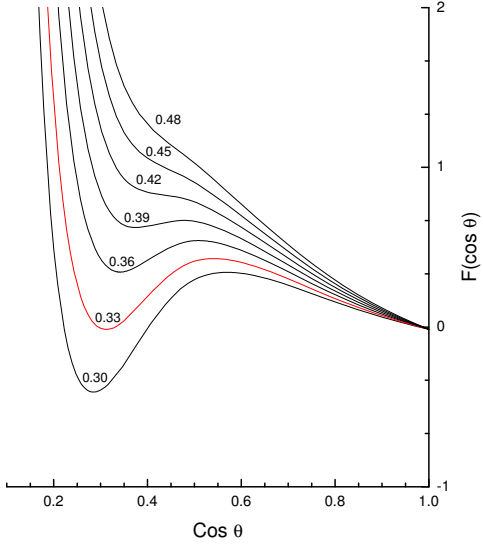}
\vskip -.5cm\caption{Dimensionless mean-field free energy $F=6{\cal
F}/W$ as a function of $\cos \theta $ for $J_H\,s/W =0.95$, $\mu/W
=-5$, and $W/U$=0.30; 0.33; 0.36; 0.39; 0.42; 0.45;
0.48}\label{fig5}
\end{center}
\end{figure}
\begin{figure}[!ht]
\begin{center}
\vspace{-0.cm} \epsfxsize=7cm \hspace*{0.2cm}
\epsfbox{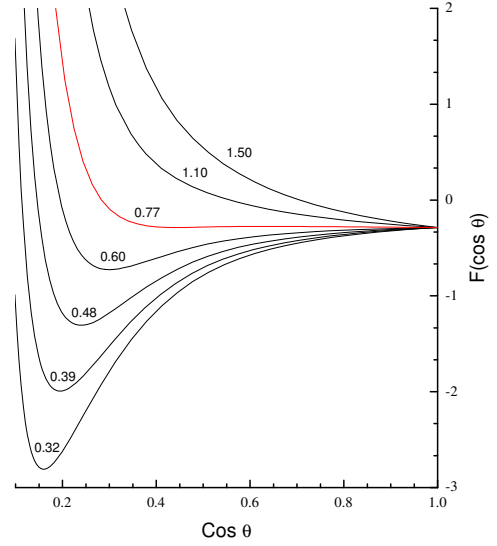}\vskip -0.3cm \caption{Dimensionless
mean-field free energy $F=6{\cal F}/W$ as a function of $\cos \theta
$ for $J_H\,s/W = 0.50 $, $\mu/W =-0.33$, and $W/U$=0.32; 0.39;
0.48; 0.60; 0.77; 1.10; 1.50 }\label{fig6}
\end{center}
\end{figure}

\section{\bf Conclusions}

 We have argued that the
on-site Coulomb repulsion strongly affected the magnetic
properties of spin-fermion systems. In particular, when Coulomb
repulsion is strong enough, the on-site localized and carriers'
spins become misaligned (on-site canted ferromagnetic state). As
follows from (Eq.\ref{cfm47}), $\cos  \Theta  > 0$ when $J_N  > 0$
(see Fig. 2a), and $\cos  \Theta   < 0$ when $J_N  < 0$ (Fig. 2b),
To obtain this result a double exchange model with Hubbard term
was considered. We represented the localized spins by means of
Schwinger-bosons and introduced two spin-singlet Fermion
operators. In terms of the new Fermi fields the on-site Hund's
interaction is in a diagonal form and the true magnons of the
system can be recognized. Written in terms of Schwinger-bosons and
spin-singlet fermions the theory is $U(1)$ gauge invariant. We
have shown that on-site canted ferromagnetic state is a state with
spontaneously broken gauge symmetry. This is because the order
parameter is gauge varying collective field with charge -2  (see
equations (\ref{cfm10}) and (\ref{cfm35})) \be\label{cfm48}
\Delta'_j(\tau) = e^{-i 2 \alpha_j(\tau)}\Delta_j(\tau) \ee and
non-zero expectation value $ \langle \Delta_j(\tau) \rangle \neq
0$ means spontaneous breakdown of the gauge symmetry.

To study the Goldstone modes in the on-site canted ferromagnetic
phase it is convenient to represent the Schwinger-bosons and the
order parameter in the form \bea\label{cfm49} & & \varphi_{j1} =
e^{i\phi_j} \sqrt {2s-\frac sM a^+_ja_j}, \qquad \varphi_{j2} =
\sqrt {\frac sM}\,\,\, a_j\,\,,\nonumber \\ & & \Delta_j =
|\Delta_j|e^{i \chi_j}. \eea
 Then the new fields, $a^+_j,\,a_j,\,\phi_j$ and
$ \chi _j$ transform under the gauge transformation in the following
way \be\label{cfm50} a'_j = e^{i\alpha_j}a_j, \,\,\,\phi'_j = \phi_j
+ \alpha_j,\,\, \, \chi '_j =  \chi _j - 2\alpha_j. \ee The theory
is an $U(1)$ gauge theory and we have to impose one gauge fixing
condition. Hence, there are two Goldstone modes in the theory. For
example one can use $\phi_i = 0$ as a gauge fixing condition (see
equation (\ref{cfm21})), then the Goldstone modes are the magnons
$a_i\,(a^+_i)$ and $ \chi _i$ phase. The physical origin of the
extra mode is the totaly broken rotation symmetry, while
mathematical reason is the spontaneous breakdown of the gauge
symmetry. Alternatively, one can choose $ \chi _i = 0$ for the gauge
fixing. Both these conditions are not convenient. In the quadratic
parts of the corresponding effective theories there are terms which
mix magnons and phases, These terms are obstacle to recognize the
spectrum in the theory. One hopes that there is a gauge fixing
condition which involves all gauge varying fields and the quadratic
terms of the effective theory takes diagonal form. This issue will
be addressed elsewhere.


\begin{thebibliography}{99}
%
\bibitem[*]{byline} Electronic address: naoum@phys.uni-sofia.bg

\bibitem{cfm01} C. Zener, Phys. Rev. {\bf 81}, 440 (1951).
\bibitem{cfm02} T. Kasuya, Prog. Theor. Phys. {\bf 16}, 45 (1956).
\bibitem{cfm021}K. Yosida, Phys. Rev., {\bf 106}, 893 (1956).
\bibitem{cfm03} J. Kondo, Prog. Theor. Phys. {\bf 32}, 37 (1964).
\bibitem{cfm04} P. Prelov$\check{s}$ek, Phys.Lett. {\bf A126}, 287 (1988).
\bibitem{cfm05} E. L. Nagaev, Phys.Status.solidi (b), {\bf 65}, 11 (1974).
\bibitem{cfm1} Vitor M. Pereira, J. M. B. Lopes dos Santos, Eduardo V.
Castro, and A. H. Castro Neto, Phys. Rev. Lett. {\bf 93}, 147202
(2004).
\bibitem{cfm11} S. S\"ullow et al., Phys.Rev {\bf B57}, 5860 (1998).
\bibitem{cfm12} S. Paschenl et all,  Phys. Rev. B {\bf 61},
4174 (200).
\bibitem{cfm13} W. Henggeler et all, Solid State Commun. {\bf 108},
929 (1998).
\bibitem{cfm20} H. Ohno, et all, Appl.Phys.Lett. {\bf 69}, 363
(1996).
\bibitem{cfm21} A. Chattopadhyay, S. Das Sarma, and A. J. Millis,
Phys.Rev. Lett., {\bf 87}, 227202 (2001).
\bibitem{cfm22} F. Popescu, Y. Yildirim, G. Alvarez, A. Moreo, and E.
Dagotto, Phys. Rev. {\bf B73}, 075206 (2006).
\bibitem{cfm2} E. Dagotto, \emph{Nanoscale Phase Separation and Colossal
Magnetoresistance} (Springer-Verlag, Berlin, 2003) and references
therein.
\bibitem{cfm31} S. Yunoki, J. Hu, A. L. Malvezzi, A. Moreo, N.
Furukawa, and E. Dagotto, Phys. Rev. Lett. {\bf 80}, 845 (1998).
\bibitem{cfm32} A. Chattopadhyay, A. J. Millis, and S. Das Sarma,
Phys.Rev. {\bf B64}, 012416 (2001).
\bibitem{cfm33} Daniel P. Arovas, and Francisco Guinea, Phys.Rev {\bf B58}, 9150 (1998).
\bibitem{cfm34} D. Pekker, S. Mukhopadhyay, N. Trivedi and P. Goldbart,
Phys.Rev {\bf B72}, 075118 (2005).
\bibitem{cfm35} M. Kagan, D. Khomskii, and M. Mostovoy,
Eur. Phys. J. {\bf B12}, 217 (1999).
\bibitem{cfm41} J. R. Schrieffer, and P.A. Wolf, Phys.Rev. {\bf 194}, 491 (1966).
\bibitem{cfm42} D. Meyer and W. Nolting, J. Phys.: Condens. Matter
{\bf 11}, 581 (1999).
\bibitem{cfm43} E. L. Nagaev, Phys. Rev. {\bf B58}, 827 (1998).
\bibitem{cfm51} Nic Shannon and  Andrey Chubukov, Phys. Rev. {\bf B65}, 104418 (2002).
\bibitem{cfm52} N. Karchev, arXiv: cond-mat/0308134 (2003).
\bibitem{cfm53} D. Pekker, S. Mukhopadhyay, N. Trivedi, and P. M. Goldbart, Phys. Rev. {\bf B72}, 075118 (2005).
\bibitem{cfm61} J. Inoue, and S. Maekawa, Phys. Rev. Lett. {\bf 74}, 3407 (1995).
\bibitem{cfm62} S. K. Mishra, R. Pandit, and S. Sathpathy, J. Phys.: Condens. Matter
{\bf 11}, 8561 (1999).
\bibitem{cfm63} D. I. Golosov, Phys. Rev. {\bf B71}, 014428 (2005).
\bibitem{cfm64} J. Medvedeva, V. Anisimov, O. Mryasov, and A. Freeman, J. Phys.: Condens. Matter
{\bf 14}, 4533 (2000).
\bibitem{cfm65} K. Held, and D. Vollhardt, Phys. Rev. Lett. {\bf 84}, 5168 (2000).
\bibitem{cfm71} P. -G. de Gennes, Phys.Rev., {\bf 118}, 141 (1960).
\bibitem{cfm72} N. B. Ivanov, J. Richter, and D. J. J. Farnell, Phys. Rev. {\bf B66}, 014421 (2002).
\bibitem{cfm73} S. Sachdev, and T. Senthila, Ann.Phys.(NY) {\bf 251}, 76
(1996).
\bibitem{cfm8} Naoum Karchev, Phys. Rev. {\bf B57}, 10913 (1998).

\end{thebibliography}
\end{document}